\documentclass[lnbip]{svmultln}
\usepackage{amsmath}
\usepackage{makeidx}  % allows for indexgeneration
\usepackage{orcidlink}
\usepackage{hyperref}
\usepackage{silence}
\begin{document}
\mainmatter              % start of the contribution
\title{Human-AI Interaction Requirements in Public Sector Procurements}
%
%\titlerunning{Human-AI Interaction Requirements}  % abbreviated title (for running head)
%                                     also used for the TOC unless
%                                     \toctitle is used
%
\author{Mateen A. Abbasi\inst{1}\orcidlink{0000-0001-8988-8816} \and 
%Sanni Marjanen\inst{1}\orcidlink{0009-0009-2008-4970} \and 
Tommi Mikkonen\inst{1}\orcidlink{0000-0002-8540-9918} \and
Sinna Pirinen\inst{1}\orcidlink{0009-0002-4047-3534} \and 
Aapo Koski\inst{2,3}\orcidlink{0000-0002-4000-8491}}
\authorrunning{Mateen A. Abbasi et al.}   % abbreviated author list (for running head)
%
%%%% list of authors for the TOC (use if author list has to be modified)
\tocauthor{Mateen Abbasi, Aapo Koski, Sinna Pirinen, Tommi Mikkonen}
\institute{University of Jyväskylä, Jyväskylä, Finland,\\
\email{{mateen.a.abbasi, tommi.j.mikkonen}@jyu.fi},\\
\email{sinna.e.pirinen@student.jyu.fi},\\
\and Tampere University, Tampere, Finland,
\email{aapo.koski@tuni.fi},\\
\and 61N Solutions Oy, Tampere, Finland,
\email{aapo.koski@61n.fi}}

\maketitle              % typeset the title of the contribution

\begin{abstract}        % give a summary of your paper
Public sector organizations increasingly procure AI-enabled ICT systems to support decision-making and service delivery. Although ethical AI frameworks emphasize transparency, accountability, and human oversight, these principles are rarely translated into explicit requirements in procurement processes. Consequently, human–AI interaction (HAI) is often left to vendor design choices. This paper conceptualizes HAI as a procurement-critical design dimension and proposes a taxonomy of interaction requirements tailored to public sector ICT procurement. The taxonomy enables contracting authorities to specify and govern interaction properties through procurement instruments, supporting both ethical compliance and sustainable value realization.

%Public sector organizations are increasingly procuring ICT systems enabled by artificial intelligence (AI) to support decision-making, service delivery, and operational efficiency. While ethical AI principles and regulatory guidance emphasize the importance of human oversight, transparency, and accountability, these concerns are rarely translated into explicit, verifiable requirements in public procurement processes. As a result, human–AI interaction (HAI) is often left to vendor design choices, creating risks related to accountability, user trust, operational effectiveness, and ethical compliance. This paper examines human–AI interaction as a procurement-critical design dimension rather than a purely technical or ethical afterthought. Building on existing ethical AI and human–AI interaction literature, the study develops a structured taxonomy of HAI requirements tailored to the context of public sector ICT procurements. By making human–AI interaction explicitly procurable, the proposed approach supports both ethical compliance and the realization of sustainable organizational value from AI-enabled systems. The findings offer practical guidance for contracting authorities and contribute to ongoing research on responsible and value-oriented AI adoption in public sector information systems.

% keywords within the abstract
\end{abstract}
\section{Introduction}
\label{sec:intro}

% Purpose: Establish the problem, relevance, and contribution.

Public sector organizations are increasingly adopting AI-enabled ICT systems to support decision-making, service delivery, and operational efficiency. Such systems are used in a wide range of contexts, including administrative decision support, data-driven process automation, and predictive analytics in regulated and, in some cases, safety-critical environments \cite{hickok2024public}. Unlike in some other contexts, the procurement of ICT systems in the public sector takes place under strict legal and ethical requirements of transparency, lack of discrimination, and clear assignment of liability. All these aspects make AI system procurements a challenge for governance rather than a technical exercise.

In this study, AI refers broadly to computational techniques that generate predictions, classifications, or recommendations from data, including machine learning models and newer generative approaches such as large language models.

One of the most important issues is how human interaction with the AI system takes place. Ethical AI frameworks, regulatory guidance, and policy initiatives consistently emphasize human oversight, explainability, accountability, and user trust \cite{nagitta2022human}. However, these principles are typically articulated at a high level and implicitly assume that appropriate interaction mechanisms will be addressed during system design and implementation. In practice, public sector organizations rely primarily on procurement processes to define system requirements and govern supplier behavior \cite{andersson2025assessing}. If human-AI interaction (HAI) is specified in the procurement procedure at all, decisions regarding how the system should act are left in the hands of the suppliers. This significantly reduces the ability of the contracting authority to ensure that the principles are followed. In this regard, the issue of human interaction with AI-based systems in the public sector is viewed from the aspect of usability. This is inadequate since the behavior of the systems is based on probability. In public sector contexts, where AI-supported decisions may carry legal, societal, or safety-critical consequences, this gap poses significant governance risks. The emergence of regulatory initiatives such as the EU AI Act further highlights the need to operationalize oversight, transparency, and accountability as verifiable requirements rather than aspirational principles.

This paper argues that HAI must be considered a procurement-critical design dimension rather than an add-on in technical or ethical considerations. Based on previous research on ethical AI, human–AI interaction, and public sector ICT procurement, this paper proposes a structured taxonomy of HAI requirements tailored to public procurement contexts. The taxonomy identifies key interaction dimensions that can be explicitly specified, evaluated, and governed through procurement tools. By linking ethical AI principles with procurable and verifiable interaction requirements, the study contributes to responsible, value-oriented AI adoption in public sector information systems and offers practical guidance for contracting authorities to balance innovation, accountability, and sustainable value realization.

\vspace{-3mm}
\section{Background and Related Work}
\label{sec:background}
% Purpose: Ground the work in existing literature, without overloading.

Research on AI in the public sector, human–AI interaction (HAI), and ethical AI provides important conceptual foundations, yet their integration into procurement practice remains limited. Existing HAI and ethical AI research focuses primarily on system design and development contexts, whereas public sector organizations typically influence system behavior through procurement requirements.

Public sector ICT procurement operates under strict legal and accountability constraints that shape verification, liability, and long-term operation \cite{ghezzi2023public}. The introduction of AI amplifies existing challenges such as requirement uncertainty and information asymmetry while introducing additional concerns related to opacity, probabilistic outputs, and loss of control \cite{andersson2025assessing,hickok2024public}. 

HAI research examines how humans interpret, supervise, and collaborate with AI systems, emphasizing explainability, trust, and levels of automation \cite{abbasi2025towards,amershi2019guidelines}. Ethical and regulatory frameworks similarly stress principles such as transparency, human oversight, and accountability \cite{cavalcante2023meaningful,kazim2021high}. However, these principles are rarely translated into concrete requirements that can be specified in tender documents, evaluated during supplier selection, and verified during system deployment \cite{perez2023trustworthy}. 

This paper addresses this gap by conceptualizing HAI as a procurement-critical design dimension and by proposing a taxonomy of interaction requirements that can be explicitly specified, evaluated, and governed through public sector ICT procurements.

\vspace{-3mm}
\section{Research Approach}
\label{sec:research-approach}

This study adopts a conceptual design approach to develop a taxonomy of human–AI interaction (HAI) requirements for public sector ICT procurement. The taxonomy was derived through a literature synthesis across four domains: HAI research, ethical and responsible AI, requirements engineering, and public sector procurement.

Recurring themes across these domains—such as transparency, human oversight, accountability, user competence, and system adaptivity—were identified and consolidated into analytically distinct interaction requirement dimensions. Particular attention was paid to ensuring that the resulting dimensions correspond to properties that can be explicitly specified and evaluated within procurement processes.

To ensure practical relevance, the taxonomy was aligned with common public procurement lifecycle stages including needs assessment, tender specification, supplier evaluation, contracting, acceptance testing, and operational monitoring \cite{rolfstam2013public}. The contribution of the study is conceptual; empirical validation is left for future work.

\vspace{-3mm}
\section{Requirement Taxonomy for Public Sector Procurement}
\label{sec:hai-interaction-req}

% This is our core contribution.
% - Introduce the idea that HAI requirements must be explicitly procurable.

This section presents a taxonomy of HAI requirements that are explicitly formulated to be procurable in public sector ICT acquisitions. Existing ethical guidelines and regulatory frameworks emphasize human oversight, transparency, and accountability, but rarely translate into concrete requirements that can be specified, evaluated, and enforced through public procurement processes \cite{johnson2025legacy}. As a result, human–AI interaction is often ambiguous or left for AI vendors to implement.

The proposed taxonomy addresses the aforementioned gap by conceptualizing human–AI interaction not as a secondary usability or ethical concern but as a specific set of interaction characteristics that must be specified clearly in procurement documents. These characteristics define how AI systems are used and monitored and can be managed through the specifications and acceptance tests of the procurement. The taxonomy gives the procuring authority the capacity to influence the interaction design and accountability structures before systems are deployed.

The taxonomy comprises five dimensions: (1) transparency and explainability, (2) human oversight and control, (3) user competence and cognitive load, (4) accountability and traceability, and (5) adaptivity and behavioral predictability. Together, these dimensions capture the core ways in which humans and AI systems interact in public sector operational contexts.

%Building on our earlier work on human–AI collaboration in requirements engineering as presented in HARE-SM framework \cite{abbasi2025towards}, this taxonomy generalizes existing process towards procurement and governance perspective. While HARE-SM focused on structuring collaboration between human stakeholders and AI techniques during requirements engineering phases, this taxonomy views human-AI collaboration not from the development view of how to collaboratively work together but from the procurement view regarding requirements that can be explicitly procurable as system usage requirements after system deployment.

\begin{figure}[ht]
\centering
\includegraphics[width=\textwidth]{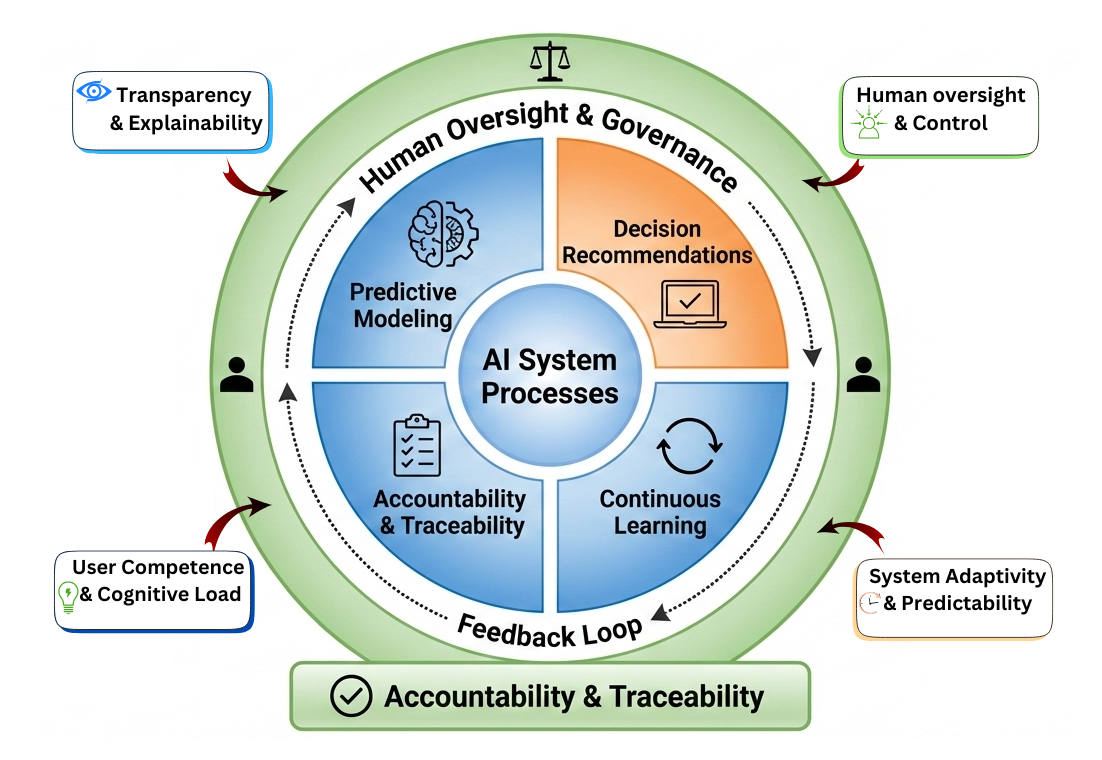}
\caption{Human–AI interaction requirements as a governance layer around AI system processes in public sector deployments}
\label{fig:HAI}
\end{figure}

Figure~\ref{fig:HAI} presents an overview of this conceptual framework. It identifies the requirements of human--AI interaction as a separate governance layer that encircles and challenges the operations of AI-enabled systems. This structure not only specifies the requirements of human and AI interaction, but also emphasizes how interaction requirements mediate between ethical and legal expectations, organizational accountability, as well as the operations of AI systems.

%As illustrated in Figure~\ref{fig:HAI}, the five dimensions of the interaction requirements condition the interpretation, validation, overriding, and adaptation of AI system outputs. Transparency and explainability determines the degree to which AI system outputs are interpretable and validatable; Human oversight and control defines the limits and effectiveness of human intervention; User competence and cognitive load assesses the degree to which competent public sector actors are able to make effective use of the systems; Accountability and traceability enable auditing, justification, and responsibility allocation; Adaptivity and behavioral predictability defines the limits to which systems adapt after implementation. The feedback loop highlighted in the figure underscores that these interaction requirements are not static, but continuously informed by system use, oversight activities, and organizational learning.

The remaining part of this section discusses each of the five dimensions, outlining their conceptual rationales, typical procurement-ready requirement formulations, and relevance for public-sector governance and accountability.

\subsubsection{Transparency \& Explainability}

Transparency and explainability refer to the extent to which AI system outputs and reasoning processes are understandable to human users. In public sector contexts this is essential for legal accountability and the justification of decisions affecting citizens \cite{wachter2017right}.

From a procurement perspective, transparency and explainability have to be formulated in concrete terms regarding interaction, rather than being considered as abstract system qualities. This involves specifying

\begin{itemize}
    \item what aspects of system behavior must be explainable (e.g., individual decisions, model logic, data influences),
    \item to whom explanations must be available (e.g., end users, supervisors, auditors),
    \item at what stages explanations are required (e.g., in real time, post hoc, during audits).
\end{itemize}

\noindent
Insufficiently specified transparency requirements risk leaving explainability as a best-effort feature rather than a guaranteed capability. Explicit procurement requirements can define minimum levels of explanation, acceptable forms of explanation, and verification methods during acceptance testing, thereby embedding transparency into contractual obligations.

\subsubsection{Human Oversight \& Control}

Human oversight and control address how and under what conditions humans can supervise, intervene, or override AI-enabled processes. This dimension formalizes and operationalizes a well-established rule of keeping a human “in the loop” or “on the loop” \cite{kazim2021high,perez2023trustworthy}.
In terms of procurement perspective, interaction mechanisms must be specified rather than assumed. The following requirements must be defined:

\begin{itemize}
    \item decision points where human confirmation or authorization may be required,
    \item methods for overriding, suspending, or disabling automated recommendations,
    \item escalation paths for uncertainty, conflict, or unusual system behavior, and
    \item role-based control rights aligned with organizational responsibilities and obligations.
\end{itemize}

\noindent
Without explicit procurement requirements, oversight mechanisms may be implemented in ways that are technically present but practically unusable, for example, due to poor interface design or excessive time pressure. Oversight as a key procurement requirement can ensure that control is not merely nominal but functional and aligned with the accountability structures within public sector organizations.

\subsubsection{User Competence \& Cognitive Load}

This dimension concerns the relationship between AI system behaviors with human cognitive capabilities, with an emphasis on understanding, attention, and the effort required to effectively engage with an AI system. In the public sector, there is considerable variation in levels of competence and workload from experts to generalist civil servants.
In procurement context, the issues of user competency and cognitive loads are dealt with through such explicit requirements as:

\begin{itemize}
    \item user training, on-boarding, and certification,
    \item clarity of feedback, suggestions, warnings given through the system, and
    \item constraints  on cognitive burden in time-critical situations.
\end{itemize}

\noindent
From HAI perspective, poorly designed interaction can lead to automation bias, where users over-trust system outputs, or to underutilization, where users ignore AI support due to lack of understanding or confidence. By specifying competence and workload-related requirements, contracting authorities can shape systems that support calibrated trust and sustainable use.

\subsubsection{Accountability \& Traceability}

Accountability and traceability refer to the ability to reconstruct, audit, and attribute AI-supported decisions and actions to human and system actors\cite{raji2020closing}. In the public sector, this dimension is essential for legal compliance, organizational governance, and public trust.
Procurement-oriented requirements in this dimension may address:

\begin{itemize}
    \item logging of inputs, outputs, and decision rationales,
    \item tracing system states, configurations, or model versions,
    \item attribution of actions to human roles and system components, and
    \item accessibility of records for audits and reviews.
\end{itemize}

\noindent
By integrating traceability requirements into procurement contracts, contracting authorities can ensure that accountability depends on verifiable system capabilities throughout the system life cycle, not on vendor-allocated tools or discretion.

\subsubsection{Adaptivity and Behavioral Predictability}

AI systems may change their behavior over time due to learning, data drift, or model updates \cite{cavalcante2023meaningful}. Although such adaptivity can offer performance improvements \cite{amershi2019guidelines}, for example, an increase in prediction accuracy, improved alignment with the operational context, or a decrease in error rates for specific sub-populations. Adaptivity introduces uncertainty about future system behavior, which is particularly problematic in regulated and accountability-driven public sector contexts \cite{fischer2024bridging}.

In the context of the regulated public sector that emphasizes accountability, this level of uncertainty can pose a challenge to consistency, predictability, and equal treatment. The dynamic nature of system behaviors that are not governed in a straightforward manner might leave the users puzzled regarding the reasons behind the differential treatment of similar instances.

This dimension captures HAI requirements related to:

\begin{itemize}
    \item constraints on autonomous learning and adaptation,
    \item mechanisms to notify users about changes in behavior,
    \item approval processes for model updates, and
    \item guarantees of  behavioral stability for critical system functions.
\end{itemize}

From a procurement perspective, system adaptivity must be explicitly governed to balance innovation with predictability. It is crucial to define rules of change and how changes should be introduced and documented and accepted. Thus, a contracting authority is capable of gaining control over the adaptive AI.

\subsubsection{Taxonomy as a Procurement-Oriented Design Construct}

% Emphasize: If a requirement cannot be specified, evaluated, and accepted, it cannot be governed.

Although analytically distinct, the five dimensions are interdependent in practice: for example, human oversight presupposes transparency, and accountability requires traceability and control roles. The taxonomy is intended not as a checklist of features but as a procurement-oriented construct that enables interaction requirements to be specified, evaluated, and verified. Its value lies in supporting contracting authorities in identifying interaction-related risks and expectations and translating them into procurable requirement categories. This framing positions HAI as a locus of governance rather than solely a design concern.

\vspace{-3mm}
\section{Operationalizing HAI Requirements} %in Public Procurements}
\label{sec:operationalizing-hai}

% This is where reviewers should see practical value.

While the taxonomy presented in Section \ref{sec:hai-interaction-req} structures key dimensions of human–AI interaction, its practical value depends on whether these dimensions can be embedded into concrete public procurement practices. This section demonstrates how human–AI interaction requirements can be operationalized across the procurement life cycle and translated into instruments that contracting authorities already use in practice.
The central argument is that human–AI interaction must be treated as an explicit object of procurement, subject to specification, evaluation, contracting, and verification. A set of interaction expectations should be formed through these processes to support governance, enforcement, and adaptation in time.

\subsection{From Ethical Principles to Procurement Language}

% Show how abstract principles translate into:
% - Mandatory requirements
% - Evaluation criteria
% - Acceptance conditions

The principles that must be considered are transparency, accountability, and human agency. However, these principles are not directly actionable within procurement processes. To become operational, they must be translated into interaction-oriented requirements that can be expressed in tender documents, assessed during supplier selection, and verified before deployment.
Across public procurement practice, this translation typically follows a recurring pattern:

\vspace{-2mm}
\begin{equation}
  \textit{Ethical principle} \to \textit{HAI requirement} \to \textit{Procurement clause}
\end{equation}

\noindent
For example, the ethical principle of transparency may be translated into a requirement that the system provides user-friendly explanations regarding the recommendations supported by AI. Such a requirement will then be formulated as a mandatory functional requirement and will be tested for acceptance in scenario-based acceptance tests. In a similar way, there may emerge requirements regarding override and confirmation steps for the principle of human agency. Across the five taxonomy dimensions, such translations commonly result in three types of procurement clauses:

\begin{itemize}
    \item \textbf{Mandatory requirements}, defining minimum acceptable interaction capabilities (e.g., availability of override functions, logging mechanisms, explanation interfaces).
    \item \textbf{Evaluation criteria}, differentiating competing solutions based on the quality of interaction design (e.g., usability of explanations, support for calibrated trust, clarity of system feedback).
    \item \textbf{Acceptance conditions}, specifying how interaction-related properties will be verified prior to deployment (e.g. scenario-based testing with end users).
\end{itemize}

\noindent
By structuring ethical aspirations in this way, human–AI interaction concerns become part of the formal decision logic of procurement rather than remaining implicit expectations or post hoc evaluation criteria.

For example, a procurement clause in a public case-handling system might require that every AI-generated recommendation can be overridden by a human operator and that both the recommendation and the final decision are logged for audit purposes.

\subsection{HAI Requirements Across the Procurement Life Cycle}

% Discuss how HAI must be addressed in:
% - Needs assessment
% - Tender documentation
% - Vendor evaluation
% - Contracting
% - Acceptance testing
% - Operational monitoring

Rules on HAI have to be understandable in any procurement process. Based on common public procurement process models, six key phases can be identified.

\vspace{-1mm}
\begin{enumerate}
    \item \textbf{Needs assessment:} Identify where human oversight, explanation, and accountability must be preserved before automation decisions are made.
    \item \textbf{Tender specification:} Translate interaction expectations into requirement clauses defining explanation capabilities, override mechanisms, and traceability obligations.
    \item \textbf{Supplier evaluation:} Assess interaction design through demonstrations, prototypes, or interaction plans provided by suppliers.
    \item \textbf{Contracting:} Define contractual obligations related to explanations, logging, and model updates affecting interaction behavior.
    \item \textbf{Acceptance testing:} Verify interaction requirements through scenario-based testing with representative users.
    \item \textbf{Operational monitoring:} Audit interaction behavior during system operation and manage updates affecting user interaction.
\end{enumerate}

\noindent
Addressing HAI across these phases reveals that interaction quality is not just an initial design problem, but a sustained governance concern.

\vspace{-2mm}
\subsection{Risks of Omission}

% What happens when HAI is not explicitly procured:
% - Shadow decision-making
% - Loss of human authority
% - Ethical responsibility gaps

If HAI is not addressed during procurement, several risks emerge. AI outputs may de facto determine decisions, reducing humans to passive verifiers. Human authority may erode if users cannot interpret or override system recommendations. Traceability and accountability suffer when explanation and role assignment are not contractually enforced, complicating audits and dispute resolution. Finally, adoption and value realization decline when users lack trust, understanding, or a sense of responsibility.

\vspace{-3mm}
\section{Discussion and Conclusions}

This study positions human–AI interaction as a procurement-critical design dimension rather than solely a usability or ethical concern. Treating interaction requirements as procurable shifts part of the control over AI system behavior from vendors to contracting authorities and enables governance over transparency, oversight, and accountability throughout the system life-cycle.

The proposed taxonomy provides a structured way to translate ethical and regulatory principles into procurement instruments such as requirement clauses, evaluation criteria, and acceptance tests \cite{koski2024ethical}. In this sense, procurement functions as an important governance mechanism for responsible AI deployment.

\vspace{-3mm}
\section*{Acknowledgments}

This work has been supported by Software Engineering Doctoral Pilot, funded by the Ministry of Education and Culture, Finland, and Business Finland (project ANSE, 1822/31/2025).

% ---- Bibliography ----
%
\bibliographystyle{splncs04}
\bibliography{hai}

@inproceedings{koski2024ethical,
  title={The Ethical Landscape in Public Procurement of ICT Systems},
  author={Koski, Aapo and Pirinen, Sinna and Mikkonen, Tommi},
  booktitle={International Conference on Software Business},
  pages={35--43},
  year={2024},
  organization={Springer}
}

@article{hickok2024public,
  title={Public procurement of artificial intelligence systems: new risks and future proofing},
  author={Hickok, Merve},
  journal={AI \& society},
  volume={39},
  number={3},
  pages={1213--1227},
  year={2024},
  publisher={Springer}
}

@article{nagitta2022human,
  title={Human-centered artificial intelligence for the public sector: The gate keeping role of the public procurement professional},
  author={Nagitta, Pross Oluka and Mugurusi, Godfrey and Obicci, Peter Adoko and Awuor, Emmanuel},
  journal={Procedia Computer Science},
  volume={200},
  pages={1084--1092},
  year={2022},
  publisher={Elsevier}
}

@article{andersson2025assessing,
  title={Assessing the value of artificial intelligence (AI) in governmental public procurement},
  author={Andersson, Per Erik and Arbin, Katarina and Rosenqvist, Christopher},
  journal={Journal of Public Procurement},
  volume={25},
  number={1},
  pages={120--139},
  year={2025},
  publisher={Emerald Publishing Limited}
}

@inproceedings{abbasi2025towards,
  title={Towards Human-AI Synergy in Requirements Engineering: A Framework and Preliminary Study},
  author={Abbasi, Mateen Ahmed and Ihantola, Petri and Mikkonen, Tommi and M{\"a}kitalo, Niko},
  booktitle={2025 Sixth International Conference on Intelligent Data Science Technologies and Applications (IDSTA)},
  pages={81--88},
  year={2025},
  organization={IEEE}
}

@inproceedings{johnson2025legacy,
  title={Legacy Procurement Practices Shape How US Cities Govern AI: Understanding Government Employees' Practices, Challenges, and Needs},
  author={Johnson, Nari and Silva, Elise and Leon, Harrison and Eslami, Motahhare and Schwanke, Beth and Dotan, Ravit and Heidari, Hoda},
  booktitle={Proceedings of the 2025 ACM Conference on Fairness, Accountability, and Transparency},
  pages={772--789},
  year={2025}
}

@article{wachter2017right,
  title={Why a right to explanation of automated decision-making does not exist in the general data protection regulation},
  author={Wachter, Sandra and Mittelstadt, Brent and Floridi, Luciano},
  journal={International data privacy law},
  volume={7},
  number={2},
  pages={76--99},
  year={2017},
  publisher={Oxford University Press}
}

@book{perez2023trustworthy,
  title={Trustworthy {AI} alone is not enough},
  author={P{\'e}rez y Madrid, Aniceto and Wright, Connor},
  year={2023},
  publisher={Dykinson}
}

@article{kazim2021high,
  title={A high-level overview of AI ethics},
  author={Kazim, Emre and Koshiyama, Adriano Soares},
  journal={Patterns},
  volume={2},
  number={9},
  year={2021},
  publisher={Elsevier}
}

@inproceedings{raji2020closing,
  title={Closing the AI accountability gap: Defining an end-to-end framework for internal algorithmic auditing},
  author={Raji, Inioluwa Deborah and Smart, Andrew and White, Rebecca N and Mitchell, Margaret and Gebru, Timnit and Hutchinson, Ben and Smith-Loud, Jamila and Theron, Daniel and Barnes, Parker},
  booktitle={Proceedings of the 2020 conference on fairness, accountability, and transparency},
  pages={33--44},
  year={2020}
}

@article{cavalcante2023meaningful,
  title={Meaningful human control: actionable properties for AI system development},
  author={Cavalcante Siebert, Luciano and Lupetti, Maria Luce and Aizenberg, Evgeni and Beckers, Niek and Zgonnikov, Arkady and Veluwenkamp, Herman and Abbink, David and Giaccardi, Elisa and Houben, Geert-Jan and Jonker, Catholijn M and others},
  journal={AI and Ethics},
  volume={3},
  number={1},
  pages={241--255},
  year={2023},
  publisher={Springer}
}

@inproceedings{amershi2019guidelines,
  title={Guidelines for human-AI interaction},
  author={Amershi, Saleema and Weld, Dan and Vorvoreanu, Mihaela and Fourney, Adam and Nushi, Besmira and Collisson, Penny and Suh, Jina and Iqbal, Shamsi and Bennett, Paul N and Inkpen, Kori and others},
  booktitle={Proceedings of the 2019 chi conference on human factors in computing systems},
  pages={1--13},
  year={2019}
}

@article{fischer2024bridging,
  title={Bridging the gap: Towards an expanded toolkit for AI-driven decision-making in the public sector},
  author={Fischer-Abaigar, Unai and Kern, Christoph and Barda, Noam and Kreuter, Frauke},
  journal={Government Information Quarterly},
  volume={41},
  number={4},
  pages={101976},
  year={2024},
  publisher={Elsevier}
}

@book{rolfstam2013public,
  title={Public procurement and innovation},
  author={Rolfstam, Max},
  year={2013},
  publisher={Edward Elgar Publishing}
}

@inproceedings{ghezzi2023public,
  title={On public procurement of ICT systems: Stakeholder views and emerging tensions},
  author={Ghezzi, Reetta and Mikkonen, Tommi},
  booktitle={International Conference on Software Business},
  pages={61--76},
  year={2023},
  organization={Springer}
}
\end{document}